# INCREASING WIRELESS SENSOR NETWORKS LIFETIME WITH NEW METHOD

Mohammad Sedighimanesh[1], Javad Baqeri[2] and Ali Sedighimanesh[3]

[1,2,3]Department of Electrical, Computer and It Engineering, Islamic Azad University of Qazvin, Qom, Iran

*Abstract*

*One of the most important issues in Wireless Sensor Networks (WSNs) is severe energy restrictions. As the performance of Sensor Networks is strongly dependence to the network lifetime, researchers seek a way to use node energy supply effectively and increasing network lifetime. As a consequence, it is crucial to use routing algorithms result in decrease energy consumption and better bandwidth utilization. The purpose of this paper is to increase Wireless Sensor Networks lifetime using LEACH-algorithm. So before clustering Network environment, it is divided into two virtual layers (using distance between sensor nodes and base station) and then regarding to sensors position in each of two layers, residual energy of sensor and distance from base station is used in clustering. In this article, we compare proposed algorithm with well-known LEACH and ELEACH algorithms in homogenous environment (with equal energy for all sensors) and heterogeneous one (energy of half of sensors get doubled), also for static and dynamic situation of base station. Results show that our proposed algorithm delivers improved performance.*

## 1. INTRODUCTION

By recent improvement of technology and growing demand for easily receive information from various environments, tracking and so on, scientists have innovated Wireless Sensor Networks (WSNs). These networks consist of a large amount of small nodes with limited range of applications. These nodes which is called sensor, can sense specific features (such as humidity, temperature, pressure and so on) in the environment around them and transmit it to their neighbors. In other words, main application of these sensors is sensing specific parameters around them and connecting with other sensors to transmit this achieved data. Although it is possible in some applications that sensors get connected to each other with communication cables, but in most of cases network is totally wireless. Nodes in such networks are typically static or have limited mobility. One of the most important issues in such networks is high possibility of failure in nodes. These failures can be occurred because of various reasons, for example when sensor nodes run out of energy. So energy is regarded as a crucial factor for network. One of the most considerable topics in these networks is energy maintenance to increase network lifetime[1, 2]. The most important issue in sensor networks is routing and the most important issue in routing is optimal energy consumption in sensors to increase network lifetime, because sensors have limited energy and are not rechargeable. These networks typically have static nodes or with limited mobility and a central node which collects sensed data from nodes directly (one-step method) or indirectly (multi-steps). In directly transmission, each sensor sends information directly to central node, because of distance between sensors and base station, a lot of energy consumed in each transmission. In contrast designs which make communication distance smaller could extend network lifetime. Clustering protocols are appropriate methods for extending WSNs lifetime. In clustering, network is divided to clusters, in each cluster a node will be selected as cluster head. Member clusters send processed data to cluster head (either directly or indirectly and by multi-steps method). After that data are aggregated and be sent to base station using one-step or multi-steps transmission [3, 4].

In WSN, the synergy between the sensor nodes is important for two reasons [5, 6]:

- The data gathered by some sensor nodes can provide a valuable inference about the environment; that is because the data have been processed after being gathered, and by putting the data together, the good results will be obtained.
- Synergy between the sensor nodes can be considered as a kind of compromise between the cost of communications and the energy of calculations. That means that the sensors cooperate with each other and send the data to the central station step by step, instead of sending the information directly to the central station and consuming high amount of energy

The main purpose of hierarchical protocols (based on clustering) is using an appropriate method for optimal use of energy sources. This is done by multi-steps transmission in network and also combination of a cluster's information to reduce transmission data load. LEACH-protocol is one of the first hierarchical protocols introduced for WSNs and a lot of application protocols have been designed based on it.

These are reasons why LEACH-protocol is important for researchers[7, 8]:

- In this protocol, clustering is done randomized, adaptive and self-organizing. Here some explanations have been added to clarify each of these features. Randomized: it means that in each round, a specific number of nodes select themselves randomly as the cluster head and being cluster head has not been predetermined for specific nodes. Adaptive: it means that nodes which have been cluster head in current round could not be candidate in next round for being head cluster. So in each round, candidates of cluster head are determined according to previous round. It is expected that all of nodes could be head cluster, after a specific number of round is done. Self-organizing: it means that network nodes in this protocol, make cluster without any special node in network or even an external factor, and this way help scalability of this protocol.

- In LEACH, data transmission from nodes of a cluster to head cluster and from head cluster to base station is done using local control and doesn't need an external factor or a specific node in network to data transmission.
- MAC-protocol used in LEACH, help saving energy by relaxing sensors appropriately in time of need.
- LEACH-protocol, like other protocols based on clustering, uses combination of data in each cluster and sends compressed data to base station. So using LEACH-protocol will lead to decrease in the number of send and receive operations in network. Meanwhile, redundant data (caused by Proximity of sensors in a cluster) will be omitted before send to base station.

Hinzelmann proposed a hierarchical routing algorithm for sensor networks called LEACH [7, 9]. LEACH is one of the most popular hierarchical routing algorithms for sensor networks. It is a clustering protocol consisting of distributed data of clusters. LEACH selects some of the sensors randomly as the head cluster (CH) and distributes energy among them. The idea is that node clustering is done based on received signal power and head clusters are using as routing to sinks. As a consequence, energy will be saved because, instead of all nodes, only head clusters do transmission. LEACH is completely distributed and doesn't need information throughout the network. However, LEACH is using single hop routing in which every node can send data directly to head cluster and base station. An Optimal number for head cluster is almost 5% of whole nodes. Processing data such as data releasing and aggregating is done locally in head clusters. Head clusters change randomly to balance energy dissipation in nodes. A random number (Integer), r will be selected between 1 and 0. A node could be current round's head cluster only if its number is below the threshold value.

$$T(n) = \begin{cases} \dfrac{p}{1 - p(r \cdot \bmod \dfrac{1}{p})} & if \quad n \in G \\ 0 & \end{cases} \qquad (1)$$

Where P is desired percentage for clusters head, G is set of nodes which have not been head cluster in last round. Nodes will be paired randomly and dynamic clustering enhance network lifetime. So could not be adopted for extend networks.

According to the studies conducted on this algorithm which are presented in article[8, 9], the most important weaknesses of this algorithm include:

- It cannot be used for the vast networks
- It is not clear that how the predetermined number of the cluster-heads (p) can be distributed equally among the network. In fact there is no guarantee about the place or the number of the cluster-heads on each scenario. Therefore it would be possible that the chosen cluster-heads be centralized in some part of the network and as a result some other nodes remain with no cluster-head.

- Creation of control over the number and the place of the cluster-heads and also on the sizes of the clusters with respect to the members, has always been considered as a challenge and solving this issue requires some effective clustering algorithms in energy-consumption and therefore it balances the network's load.

Enhanced Low-Energy Adaptive Clustering Hierarchy (E-LEACH): in this protocol[10], choosing the cluster-head (CH) improves through considering the remained energy, and it is believed that the number of cluster heads is equal to the square root of the sum of the number of the sensors. In the first scenario, it is assumed that each node enjoys from equal probability of becoming a CH, but on the next scenarios they have different levels of remained energies, and on that basis they make decisions. In other words, in order to reduce the total energy consummation under specified hypotheses, the ELEACH specifies that the required number of the CHs is the square root out of the analogy of the total number of the sensor nodes. Other aspects of ELEACH are similar to the LEACH.

Designing the routing and data dissemination protocols for the WSNs is challenging due to several limitations of the network which include[11, 12]:

- The sensor's characteristics: the wireless sensor networks (WSNs) suffer from the limitations of several network sources including the energy, bandwidth, central processing unit and storing.
- The network's characteristics: the network's topology which is defined through the sensors and the communication links between the sensors, frequently changes due to increasing or deleting the sensors.
- The sensory application requirements: in most of sensory applications, the sent data shall be accurate as much as possible so that the better decision-making by the sink is assured. Furthermore, the sensed data shall fetch the sink regularly. And also, the data abundance would sometimes come desirable, when the data accuracy is increased.

The sinks on sensor networks are known as the gatherer of raw data from the sensory nodes which apply processes on raw data and deliver them to the user. In other words, they are our gates between the user and the sensory nodes which can be either mobile or fixed in sinks' sensory networks. Recent investigations found that the mobile sinks have much more benefits compared to the fixed sinks which include[5, 13]:

- The mobile sink can move throughout the sensor network, while the fixed sink cannot and is usually placed on a predefined position.
- The mobile sink results in an increase in the sensor network's lifetime and a decrease in sensor nodes energy consummation; while the fixed sink depending on where they are placed on the network cause low or high energy wastage of the sensor networks and also on the final nodes which end on the fixed sink, they result in formation of a gorge.
- The next case is the case of security on the mobile sink, because due to moving permanently, the mobile sink on sensory network can be achieved and identified by those who are not allowed to access the sink or even the network.

- Level of bearing error in sensory networks which use the fixed sink is low.

## 2. STEPS THE PROPOSED ALGORITHM

In this section, the proposed algorithm will be explained in following three steps: dividing network environment into virtual layers, clustering model, making cluster size symmetry and energy model.

### 2.1. Layering Network Environment

In our proposed algorithm, network environment is divided into two virtual layers, regarding to distance from base station which is shown in Figure 1. It is divided using Equation (2).

$$d1 = Y - y1$$
$$d2 = Y - y2$$
$$L = \frac{d1 + d2}{2}$$

(2)

Where d2 is minimum distance between sensors and base node, d1 is maximum distance between sensors and base node, and L is mean of d1 and d2. In this way network environment is divided into 2 virtual layers.

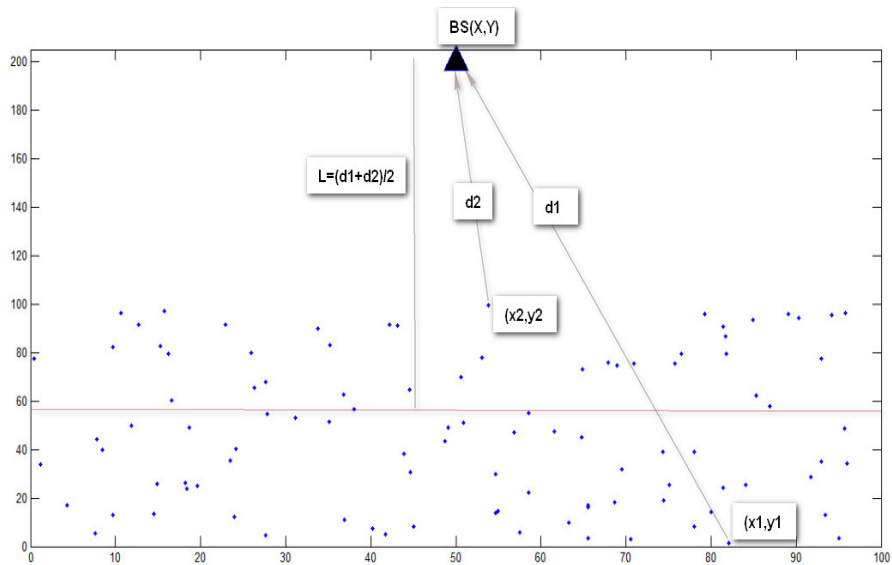

Figure 1. virtual layers of sensor network environment

### 2.2. Clustering Model

In our proposed method for clustering and cluster head selection, after dividing network environment into virtual layers using equation (2), we use equation (3) or (4) to head cluster selection process (according to sensor position in first or second layer).

If sensor is in first virtual layer, equation (3) is used for head cluster selection.

$$T(n) = \frac{p}{1 - p(r . \bmod \frac{1}{p})} \times \frac{E}{E_{in}} \times \left(\frac{d2}{D}\right)^2 \tag{3}$$

If sensor is in second virtual layer, equation (4) is used for head cluster selection.

$$T(n) = \frac{p}{1 - p(r . \bmod \frac{1}{p})} \times \frac{E}{E_{in}} \times \left(\frac{d2}{D}\right)^4 \tag{4}$$

Where E is residual energy of sensor node, $E_{in}$ is initial energy of sensor, d2 is minimum distance between sensor nodes and base station and D is distance between one sensor node and base station.

### 2.3. Making Cluster Size Symmetry

In addition, in this paper clustering is done using appropriate distribution of sensors in each cluster. According to Equation (5), after determining number of clusters head, number of clusters head (h) is subtracted from N (number of all sensors) and the result is divided to h (number of clusters head), finally floor of resulting number is considered as the number of members in each cluster (n).

$$n = \left\lceil \frac{N - h}{h} \right\rceil \tag{5}$$

In this way, number of members in each head cluster gets almost equal, therefore equal time is consumed for gathering sensed data in head clusters and also energy of head clusters is used appropriately in all clusters. As a consequence, we will have equilibrium of load and energy in each cluster and also throughout the network.

### 2.4. Model of Eenergy

Energy consumption in wireless sensor network occurs in three parts: data transmission, data reception and data processing. Model of energy is shown in equation (6) [14, 15]:

$$\begin{cases} P_T(K) = E_{elec} \times K + E_{amp} \times d^y \times K \\ P_R(K) = E_{elec} \times K \\ P_{cpu}(K) = E_{cpu} \times K \end{cases} \quad (6)$$

Where $P_T$, $P_R$, $P_{cpu}$ are energy consumption for transmission, receive and processing k bit, respectively. $E_{elec}$, $E_{amp}$, $E_{cpu}$ are per bit energy consuming (nJ/bit) for transmission in Radio range, required energy for transmission in range further than $E_{elec}$ and required per bit energy for processing, respectively. According to equation (6), total energy consumption for k bit is calculated by equation (7).

$$\begin{aligned} P_{Total} &= P_{send} + P_{Receive} + p_{cpu} \\ p_{Total} &= k(2E_{elec} + E_{cpu} + E_{amp} \times d^y) \end{aligned} \quad (7)$$

As shown above, energy consumption is directly proportional to length of data. Lower length of data, lower energy consumed. If transmission distance is lower than a threshold, energy consumed is proportional to $d^2$. If it is further than the threshold, energy consumed is proportional to $d^4$. So we can consume less energy, using lower transmission distance.

## 3. Simulation

In this section we consider our proposed algorithm, LEACH and ELEACH-algorithms in a Homogenous environment (with equal energy for all sensors) and a Heterogeneous one (energy of half of sensors get doubled). Matlab software is used for simulation. Table 1 shows the initial parameters of Wireless Sensor Networks for simulating 100 nodes.

Table 1. Initial parameters of Sensor Network

| Notation | Description |
|---|---|
| Area=100*100 | Area used in the simulation in metes |
| E0=0.5(J) | Initial energy |
| $E_{cpu}$=7(nJ/bit) | Per bit energy consuming |
| $E_{elec}$=50(nJ/bit) | Per bit energy consuming |
| $E_{amp}$=0.659(nJ/m$^2$) | Amplifier transmitting energy |
| $E_{da}$=5(nJ/bit) | Energy for data aggregation |
| Packet size | 4000 bits |

| 50*200 | Position of base station |

General assumptions in simulations were:

- The Network environment is square with defined number of sensors.
- Sensors are randomly uniformly distributed.
- Sensors are static
- Sensors are aware of their locations
- Initial energy of sensors is defined
- The sensor indication is unique
-

## 3.1 Experiment 1 and Experimental Results

In this subsection, we compare lifetime and the number of dead nodes for proposed method with LEACH and ELEACH-protocols. In this comparison, we assume that base station is static and all sensors have equal energy (Homogenous environment).

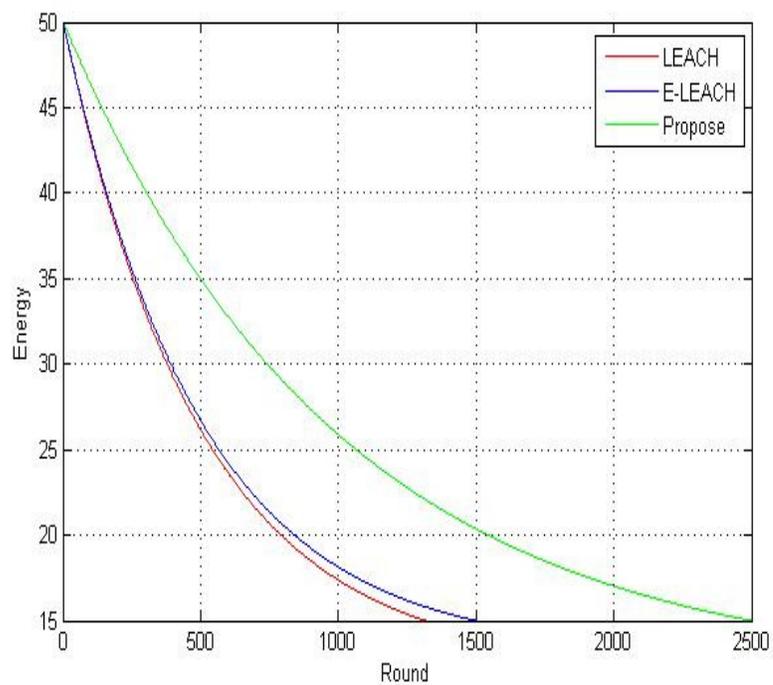

Figure 2. comparison network lifetime for proposed method with LEACH and ELEACH protocols

72

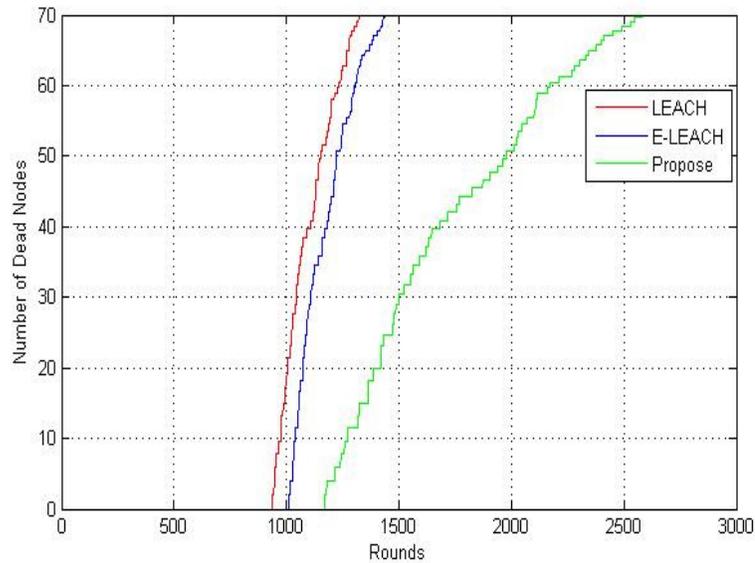

Figure 3. comparison number of dead nodes in proposed method with LEACH and ELEACH protocols

Considering assumptions of Table 1, we will compare proposed method with LEACH and ELEACH protocols in 3 different situations (death of the first node in network, 50% nodes are dead and after the number of dead nodes reaches 70%).

Table 2. Comparison lifetime for proposed methods with LEACH and ELEACH protocols, consider different percentage of dead nodes.

| death of the 70% nodes | death of the 50% nodes | death of the first node | Life time / *methods* |
|---|---|---|---|
| 2500 | 2000 | 1150 | Propose |
| 1315 | 1100 | 930 | LEACH |
| 1500 | 1200 | 1000 | ELEACH |

As shown in Figure 2, Figure 3 and also Table 2, our proposed algorithm's lifetime is more than that for LEACH and ELEACH, and performs almost 90% better than LEACH algorithm and 65% better than ELEACH algorithm.

### 3.2. Experiment 2 and Experimental Results

In this subsection comparison is done between our proposed method and LEACH, ELEACH protocols in a situation which base station is static and environment is heterogeneous (energy of half of sensors get doubled).

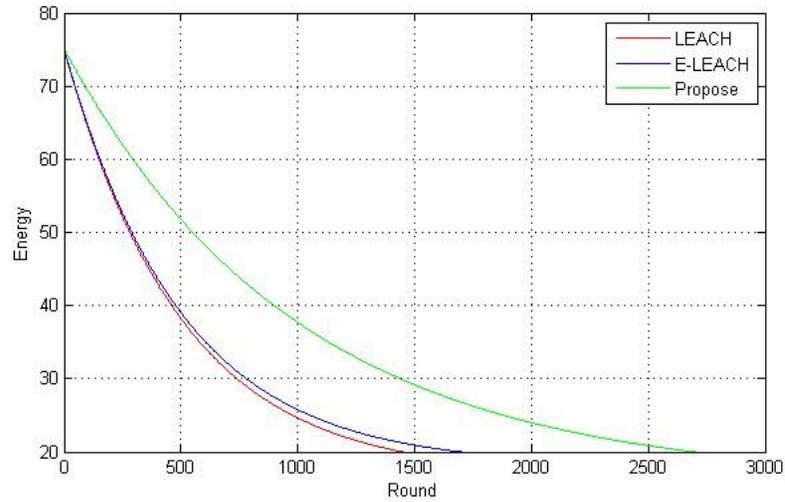

Figure 4. comparison network lifetime for proposed method with LEACH and ELEACH protocols

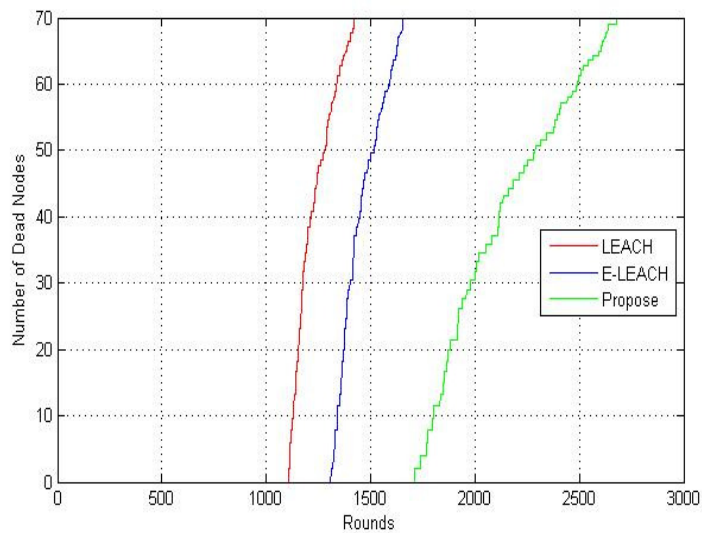

Figure 5. comparison number of dead nodes for proposed method and LEACH, ELEACH protocols

International Journal of Wireless & Mobile Networks (IJWMN) Vol. 8, No. 4, August 2016

Considering assumptions of Table 1, we will compare proposed method with LEACH and ELEACH protocols in 3 different situations (death of the first node in network, 50% nodes are dead and after the number of dead nodes reaches 70%).

Table 3. Comparison lifetime for proposed methods with LEACH and ELEACH protocols, consider different percentage of dead nodes.

| death of the 70% nodes | death of the 50% nodes | death of the first node | Life time / methods |
|---|---|---|---|
| 2700 | 2400 | 1700 | Propose |
| 1450 | 1300 | 1100 | LEACH |
| 1700 | 1400 | 1300 | ELEACH |

As shown in Figure 4, Figure 5 and also Table 3, our proposed algorithm's lifetime is more than that for LEACH and ELEACH; and performs almost 86% better than LEACH algorithm and 60% better than ELEACH algorithm.

Up to now, we assumed in our simulation that base station is static. Here we consider situation in which base station is dynamic and makes 1/10 round per second.

### 3.3. Experiment 3 and Experimental Results

In this subsection comparison is done between our proposed method and LEACH, ELEACH protocols in a situation which base station is dynamic and environment is homogenous (with equal energy for all sensors).

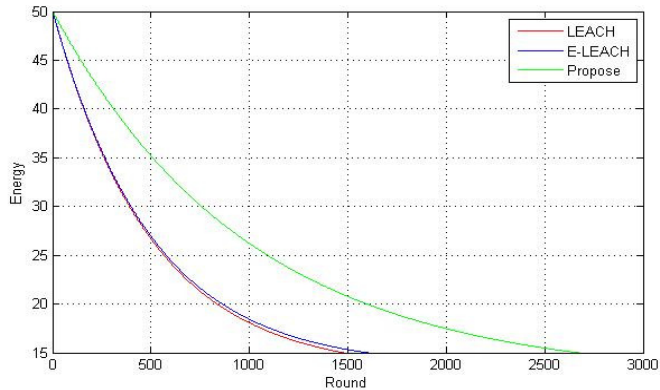

Figure 6. comparison network lifetime for proposed method with LEACH and ELEACH protocols

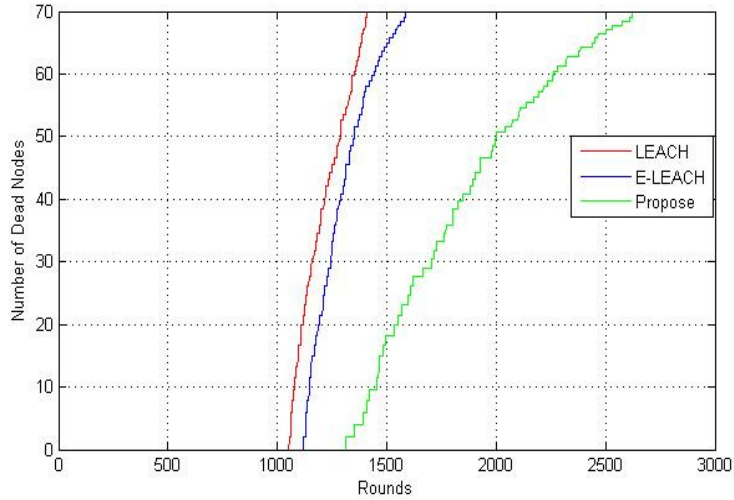

Figure 7. comparison number of dead nodes for proposed method and LEACH, ELEACH protocols

Considering assumptions of Table 1, we make comparison between proposed method with LEACH and ELEACH protocols in 3 different situations (death of the first node in network, 50% nodes are dead and after the number of dead nodes reaches 60%).

Table 4. Comparison lifetime for proposed methods with LEACH and ELEACH protocols, consider different percentage of dead nodes.

| death of the 70% nodes | death of the 50% nodes | death of the first node | Life time  methods |
|---|---|---|---|
| 2670 | 1320 | 1300 | Propose |
| 1480 | 1320 | 1050 | LEACH |
| 1600 | 1450 | 1120 | ELEACH |

As indicated in Figure 6, Figure 7 and also Table 4, our proposed algorithm's lifetime is more than that for LEACH and ELEACH; and performs almost 80% better than LEACH algorithm and 65% better than ELEACH algorithm.

## 3.2 Experiment 4 and Experimental Results

In this subsection comparison is done between our proposed method and LEACH, ELEACH protocols in a situation which base station is dynamic and environment is heterogeneous (energy of half of sensors get doubled).

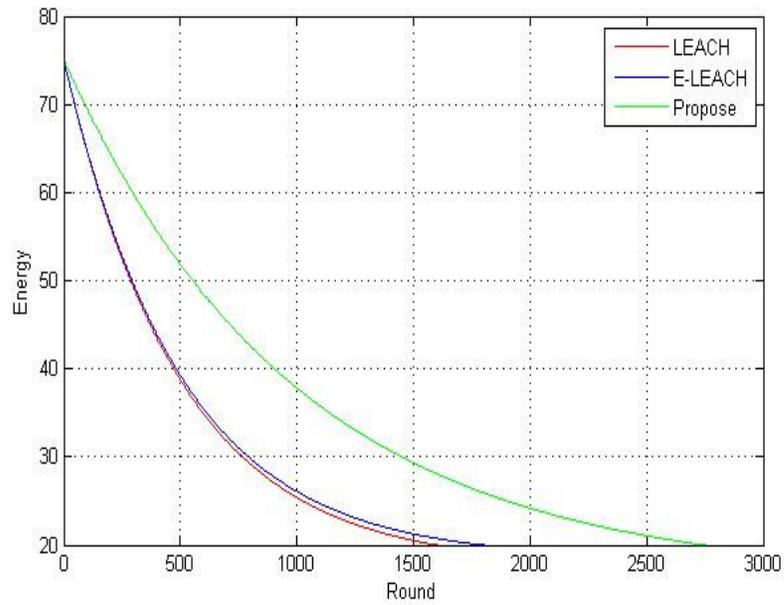

Figure 8. comparison network lifetime for proposed method with LEACH and ELEACH protocols

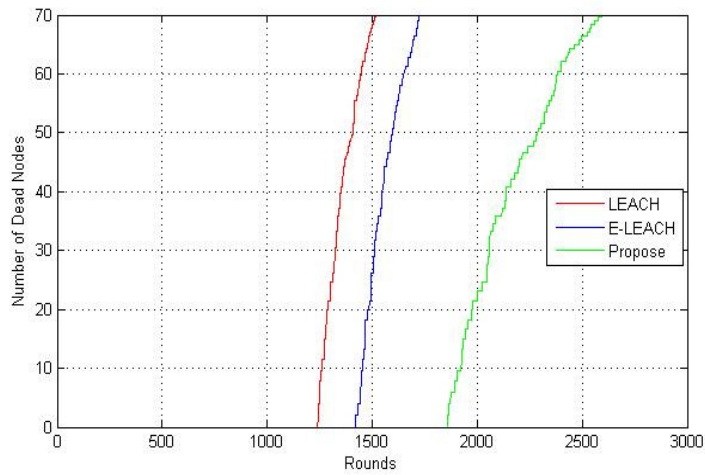

Figure 9. comparison number of dead nodes for proposed method and LEACH, ELEACH protocols

Considering assumptions of Table 1, we make comparison between proposed method with LEACH and ELEACH protocols in 3 different situations (death of the first node in network, 50% nodes are dead and after the number of dead nodes reaches 60%).

77

Table 5. Comparison lifetime for proposed methods with LEACH and ELEACH protocols, consider different percentage of dead nodes.

| death of the 70% nodes | death of the 50% nodes | death of the first node | Life time / methods |
|---|---|---|---|
| 2750 | 2400 | 1850 | Propose |
| 1600 | 1450 | 1240 | LEACH |
| 1800 | 1600 | 1420 | ELEACH |

As indicated in Figure 8, Figure 9 and also Table 5, our proposed algorithm's lifetime is more than that for LEACH and ELEACH, and performs almost 70% better LEACH algorithm and 55% better than ELEACH algorithm.

## 4. CONCLUSIONS AND FUTURE WORKS

Undoubtedly, one of the most challenging issues in Wireless Sensors Networks (WSN) is severe energy restrictions. In this paper, we described LEACH algorithm and proposed a way to improve it. Simulation results show that our suggested method is more effective than LEACH and ELEACH algorithm in WSNs whether the environment is homogeneous or Heterogeneous and whether the base station is static or dynamic.

We suggest these issues to develop proposed protocol in future works:

- Combining proposed algorithm with other routing protocols such as multi-step routing protocols.
- Adapting proposed algorithm with distributed clustering protocols.
- Applying other effective parameters for clustering;
- Considering a criterion for head cluster change time instead of changing them in each round, in this way we could save energy which is consumed for head cluster changing.
- Using a self-organizing neural cluster head.

**Authors**

**Mohammad Sedighimanesh** is a graduate student in the School of Electrical and Computer Engineering, University of Science and Technology, Qazvin Islamic azad University(QIAU),Iran. he received a Bachelor degree from University of Science and Technology Zanjan, iran. His research areas are wireless communications and Network. and a Master degree from QIAU. His current research interests include wireless and mobile communications, cooperative communications, optimization theory on communications.

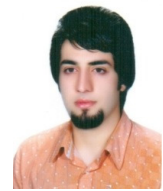

**Javad Baqeri** is a graduate student in the School of Electrical and Computer Engineering, University of Science and Technology, Qazvin Islamic azad University(QIAU),Iran. he received a Bachelor degree from University of Science and Technology Shomal, iran. His research areas are wireless communications and Network. and a Master degree from QIAU. His current research interests include wireless and mobile communications, cooperative communications, optimization theory on communications.

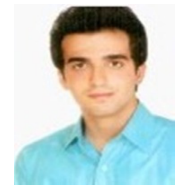

**Ali Sedighimanesh** is a graduate student in the School of Electrical and Computer Engineering, University of Science and Technology, Qazvin Islamic azad University(QIAU),Iran. he received a Bachelor degree from University of Science and Technology Parand, iran. His research areas are wireless communications and Network. and a Master degree from QIAU. His current research interests include wireless and mobile communications, cooperative communications, optimization theory on communications.

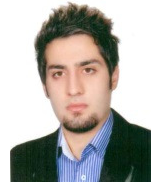